\documentclass[aps,prb,twocolumn,showpacs]{revtex4-2}

\usepackage{hyperref}
\usepackage[dvipdfmx]{graphicx}
\usepackage{amsmath}
\usepackage{amsfonts}
\usepackage{xcolor}
\usepackage{braket}
\usepackage[normalem]{ulem}
\usepackage{listings}
\hypersetup{colorlinks=true, citecolor=blue, urlcolor=blue, linkcolor=blue}

\newcommand{\s}{\mathbf{S}}

\begin{document}
\title{
Global quantum phase diagram and non-Abelian chiral spin liquid in a spin-3/2 square lattice antiferromagnet
}

\author{Wei-Wei Luo$^{1,2}$, Yixuan Huang$^{3,4}$, D. N. Sheng$^5$, W. Zhu$^{1,2}$}

\affiliation{$^1$Institute of Natural Sciences, Westlake Institute for Advanced Study, Hangzhou 310024, China}
\affiliation{$^2$School of Science, Westlake University, Hangzhou 310024, China}
\affiliation{$^3$Theoretical Division, Los Alamos National Laboratory, Los Alamos, New Mexico 87545, USA }
\affiliation{$^4$Center for Integrated Nanotechnologies, Los Alamos National Laboratory, Los Alamos, New Mexico 87545, USA}
\affiliation{$^5$Department of Physics and Astronomy, California State University, Northridge, California 91330, USA}

\date{\today}

\begin{abstract}

Since strong quantum fluctuations are essential for the emergence of  quantum spin liquids, 
there have been extensive exploration and identification of spin liquid candidates in  spin-$1/2$  systems, while such activities are rare in higher spin systems.  
Here we report an example of non-Abelian chiral spin liquid emerging in spin-$3/2$ Heisenberg model on a square lattice. 
By tuning Heisenberg exchange interaction and  scalar chirality interaction,  we map out a quantum phase diagram enclosing three conventional magnetic orders and a  chiral spin liquid based on  density matrix renormalization group studies. The nature of the spin liquid is identified as a long-sought bosonic version of Read-Rezayi state 
that supports non-Abelian Fibonacci anyonic statistics, identified by the ground state entanglement spectrum.
Significantly, we establish that the non-Abelian CSL emerges through the enlarged local degrees of freedom and enhanced quantum fluctuations near the classical phase boundaries of competing magnetic orders. Our 
numerical discovery of an exotic quantum spin liquid in spin-$3/2$ system
suggests  a new route for discovering fractionalized quantum phases in frustrated higher spin magnetic compounds.

\end{abstract}

\maketitle

\textit{Introduction.---}
One main theme in condensed matter physics is to search and classify various quantum states of matter. While most quantum phases can be heuristically understood in terms of the symmetry breaking paradigm, some strongly correlated states go beyond the conventional classification by local order parameter and interesting phenomena may emerge.
Quantum spin liquid (QSL)~\cite{balents2010,zhou2017,savary2017,broholm2020}, which does not form any conventional magnetic order even down to zero temperature, is such an example of exotic state that internally possesses fractionalized quasi-particles and long range quantum entanglement. Anderson~\cite{anderson1973} initially envisioned such a quantum disordered state to be realized in the frustrated triangular Heisenberg model. 
Such a possible QSL in the triangular Heisenberg model was predicted to be a gapped chiral spin liquid (CSL)~\cite{kalmeyer1987},
which breaks time reversal symmetry as a spin analogy of the fractional quantum Hall liquid~\cite{klitzing1980,tsui1982,laughlin1983}  exhibiting a topological order~\cite{wen1990}. Recently, using large-scale numerical simulations, CSLs have been unambiguously identified in local spin-$1/2$ models on kagome lattice~\cite{gong2014,he2014a,bauer2014,he2015b}, triangular lattice~\cite{gong2017,cookmeyer2021}, honeycomb lattice~\cite{hickey2016,hickey2017,huang2021}, and square lattice~\cite{nielsen2013,hickey2017}. 
The mechanism of the formation of the CSLs is attributed to the strong interplay of geometric frustration and quantum fluctuation. The examples of CSLs available so far all share some common features, i.e. they are produced in $S=1/2$ models where quantum fluctuations are strong and they are identified as bosonic Laughlin state processing Abelian topological order.

While the Abelian CSL appears to be common in these frustrated spin-1/2 systems,  
much less is understood regarding the emergence of CSL that possesses non-Abelian fractional statistics  in realistic local spin model.
Kitaev model is one remarkable example demonstrating the existence  of such a non-Abelian topological order, relevant to Kitaev materials~\cite{hermanns2018}. 
Besides the Kitaev materials, one natural place to search for such a state is in the systems with larger spin ($S>1/2$) ~\cite{greiter2014}.
So far, only a few studies on CSLs have been reported in spin $S=1$ system~\cite{greiter2009,liu2015b,changlani2015,liu2018,chen2018,huang2022,jaworowski2022}, which have identified  non-Abelian Moore-Read state~\cite{moore1991} with quasi-particles obeying  the  Ising anyonic statistics.
While  the Moore-Read  CSL   has potential to realize topological quantum computation~\cite{nayak2008}, from a practical point of view the Read-Rezayi state~\cite{read1999} that hosts the non-Abelian Fibonacci anyon has better performance in topological quantum computation under a noisy environment due to its universality in quantum computing algorithm~\cite{nayak2008,lahtinen2017,field2018,genetayjohansen2021}. Although there are rare cases of Fibonacci anyon that are proposed in exotic fractional quantum Hall state~\cite{xia2004,vaezi2014,zhu2015,ghazaryan2017,mong2017} and  Kondo anyons~\cite{komijani2020}, it is highly desired to search for the emergent non-Abelian CSL supporting  Fibonacci anyonic statistics in general and realistic large spin $S>1$ systems.  Such systems may be realized in quasi-two dimensional antiferromagets with 3d transition metal including  $\text{Ba}_{2}\text{Co}\text{Ge}_{2}\text{O}_{7}$~\cite{miyahara2011,romhanyi2012}.

In this paper, we address the central issue  whether the CSL can arise by suppressing magnetically ordered states in a higher spin system. Specifically we consider an antiferromagnetic Heisenberg model with quantum $S=3/2$ spins on the square lattice. 
Using large-scale density matrix renormalization group (DMRG) calculations~\cite{white1992,schollwock2011}, we establish a global phase diagram including three conventional magnetic orders which survive even in the classical limit, and importantly, among the phase boundaries of magnetically ordered states there exists a CSL state. This CSL is characterized by exponentially decaying spin correlation and characteristic level countings in entanglement spectrum~\cite{li2008} as a finger-print of the non-Abelian Read-Rezayi state~\cite{read1999}. 

\begin{figure}
        \begin{minipage}[t]{0.43\columnwidth}
        \vspace{0pt}
        \includegraphics[width=\textwidth]{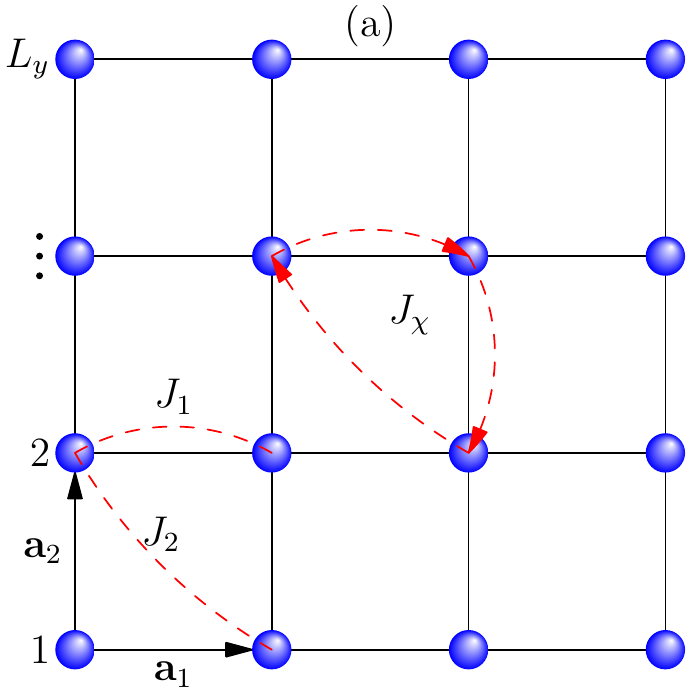}%
        \end{minipage} 
        \begin{minipage}[t]{0.5\columnwidth}
        \vspace{0pt}
            \includegraphics[width=\textwidth]{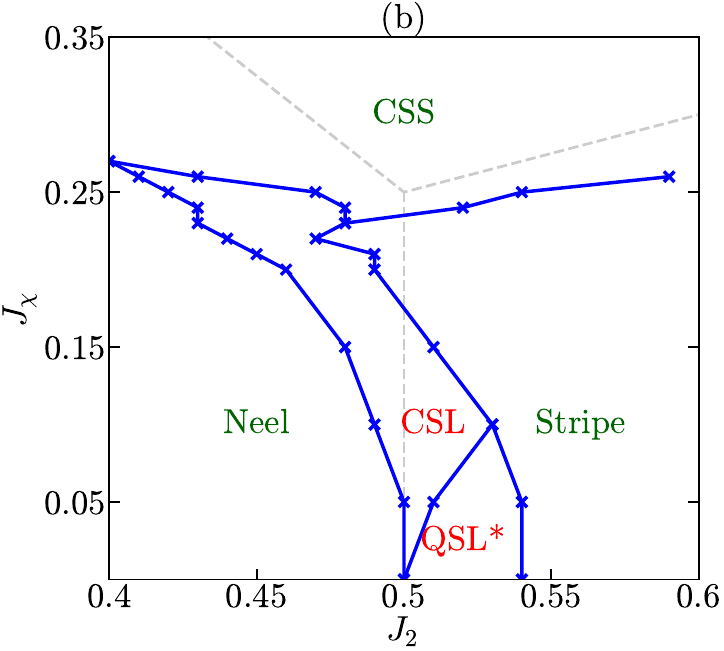}%
        \end{minipage}
\caption{(a) Spin-$3/2$ $J_1$-$J_2$-$J_\chi$ Heisenberg model on the square lattice. (b)  Quantum phase diagram by tuning  $J_2$ and $J_\chi$ (by setting $J_1=1$). CSL, CSS, and QSL* respectively denotes chiral spin liquid, chiral spin state, and possible quantum spin liquid. We also show classical phase boundaries as dashed lines, which separate three magnetic ordered phases (marked by green color).  }
\label{Fig_phase_diagram}
\end{figure}

\textit{Model and method.---}
We study the spin-$3/2$ $J_1$-$J_2$-$J_\chi$ Heisenberg model on a square lattice,
\begin{align*}
    H =& J_1\sum_{\left<ij\right>} \s_i\cdot\s_j + J_2\sum_{\left<\left<ij\right>\right>} \s_i\cdot\s_j \nonumber \\
    &+ J_\chi\sum_{ijk\in\bigtriangleup} \s_i\cdot(\s_j\times\s_k),
\end{align*}
where $\s_i$ denotes $SU(2)$ symmetric spin-$3/2$ operator on site $i$.  The exchange interactions $J_1$ and $J_2$ run over all nearest-neighbor bonds $\left<ij\right>$ and next-nearest-neighbor bonds $\left<\left<ij\right>\right>$, respectively. The three-spin scalar chiral interaction $J_\chi$ runs over all four triangles within each primitive square plaquette, and the vertices $ijk$ of each triangle are ordered in clockwise manner (see Fig. \ref{Fig_phase_diagram}(a)). This term explicitly breaks time reversal symmetry and thus favors long range chiral orders. Physically this chiral spin interaction can be deduced from the extended Hubbard model in a magnetic field, whose large repulsion $U$ limit at half filling gives rise to $J_\chi\propto  \frac{t_1^2t_2}{U^2}\sin\Phi$ for a primitive triangle enclosed by magnetic flux $\Phi$~\cite{sen1995} ($t_1$  and $t_2$ are the nearest  and next nearest neighbor  electron hoppings, respectively). In the following we fix $J_1=1$ as the unit of energy scale.

To determine possible quantum phases and quantum phase diagram, we systematically utilize both finite and infinite DMRG calculations with $U(1)$ symmetry on cylinder geometry~\cite{white1992,hauschild2018}. Due to the much larger dimension of Hilbert space compared to spin $S=1/2$ case, we mainly focus our study on finite and infinite cylinder with circumferences $L_y=4$ and $6$. We keep the bond dimension of matrix product state up to $\chi=4000$, which allows to obtain the ground state on a $L_y=4$ ($L_y=6$) cylinder with a typical truncation error of about $10^{-8}$ ($10^{-5}$).

\textit{Phase diagram.---}
Analyzing the classical spin system will give us important clues on possible quantum phases. The classical Heisenberg model on the square lattice harbors three magnetically ordered phases in $J_2$-$J_\chi$ phase diagram \cite{rabson1995}  including a Neel state at small $J_2$ regime, a stripe state at large $J_2$ regime and a chiral spin state (CSS) at larger  $J_\chi$  regime (see Fig.~\ref{Fig_phase_diagram}(b)).  
These three states meet at classical transition point $(J_2, J_\chi)=(0.5, 0.25)$, near which quantum fluctuation is expected to be strong and may promote spin disordered state in the quantum spin systems.

We present a global quantum phase diagram of spin-$3/2$ $J_1$-$J_2$-$J_\chi$ model in Fig.~\ref{Fig_phase_diagram}(b) using infinite DMRG calculations. When tuning off the chiral term $J_\chi=0$, we find the conventional Neel and stripe order at small and large $J_2$, respectively. Between them we find a quantum disordered regime near classical transition point $J_2=0.5$, where the spin-spin and dimer-dimer correlation decays exponentially. This state does not break lattice translational symmetry nor time reversal symmetry and we label it by QSL* in the phase diagram (see Supple. Mat. Sec. V \cite{SuppMaterial}). By gradually increasing $J_\chi$, both of two magnetic phases extend to a finite regime in the phase diagram. When the chiral term dominates ($J_\chi \geq 0.25$ ), a non-coplanar CSS is observed, which is also a magnetic state that survives even in the classical limit. In vicinity of classical transition boundaries, 
we discover a finite regime for CSL which hosts extremely short-ranged spin correlations. The nature of this CSL state is identified as bosonic version of Read-Rezayi state via the characteristic entanglement spectrum as we demonstrate below.

\begin{figure}
    \begin{center}
    \includegraphics[width=0.33\columnwidth]{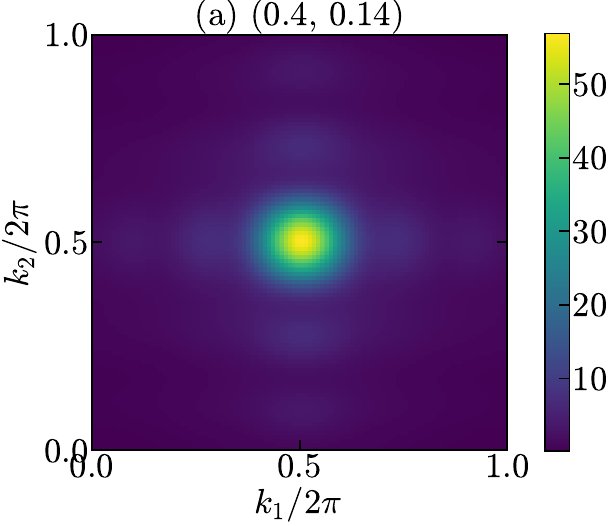}%
    \includegraphics[width=0.33\columnwidth]{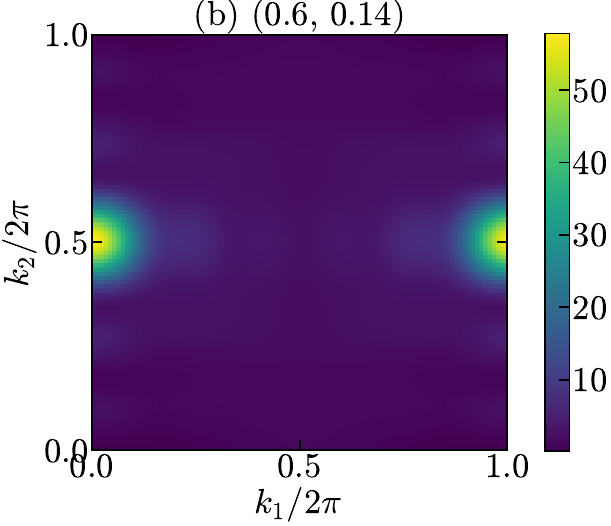}%
    \includegraphics[width=0.33\columnwidth]{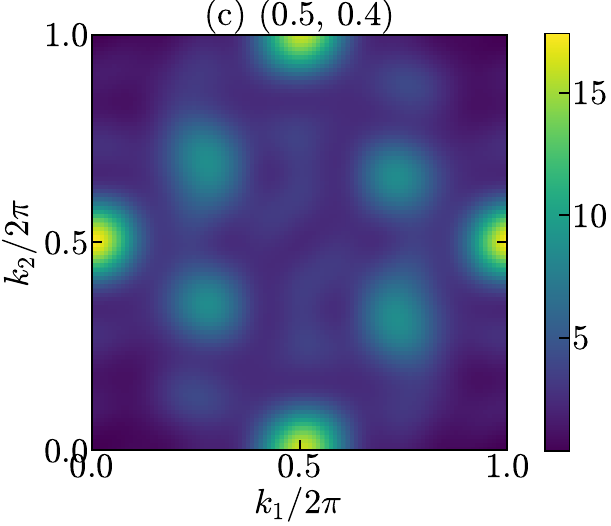}
    \includegraphics[width=0.33\columnwidth]{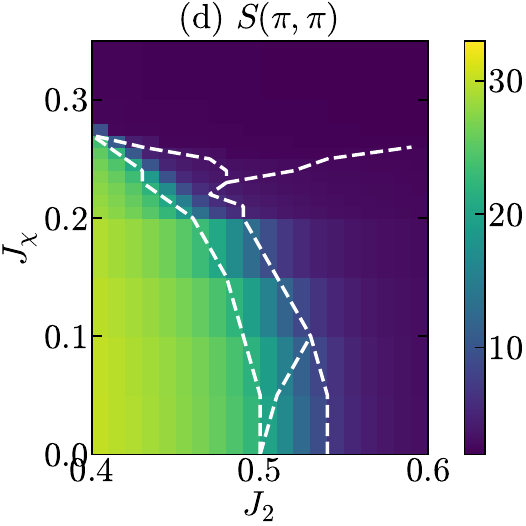}%
    \includegraphics[width=0.33\columnwidth]{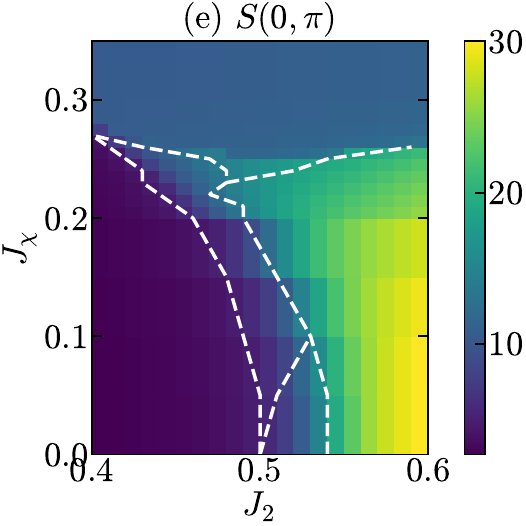}%
    \includegraphics[width=0.33\columnwidth]{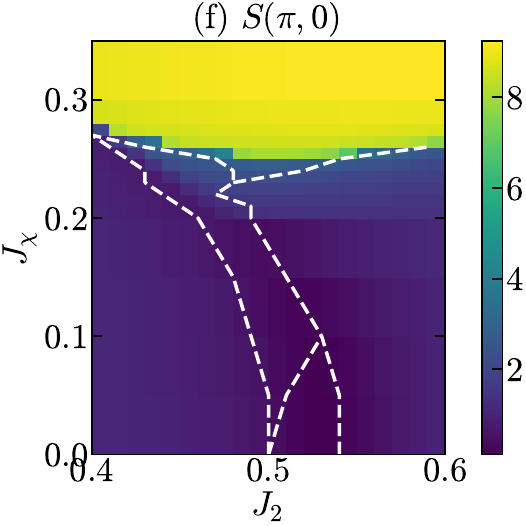}%
    \end{center}
    \caption{Upper panel: Spin structure factor $S(\mathbf{k})$ in the magnetic (a) Neel, (b) stripe, and (c) CSS phases at representative points. Lower panel: profile of magnetic order parameters (d) $S(\pi,\pi)$, (e) $S(\pi,0)$ and (f) $S(0,\pi)$ in the  $J_2$-$J_\chi$ parameter space. The white dashed line marks the phase boundary.}
    \label{Fig_structure_factor}
\end{figure}

\textit{Magnetic orders.---}
In order to determine magnetic orders, we compute spin structure factor $S(\mathbf{k})=\frac{1}{N} \sum_{i,j} \braket{\s_i\cdot\s_j}e^{i\mathbf{k}\cdot(\mathbf{r}_i-\mathbf{r}_j)} $, where both $i$ and $j$ sum over $N$ lattice sites. The structure factors in different phases are shown in the upper panel of Fig.~\ref{Fig_structure_factor}, where different peak locations specifically correspond to different magnetic orders. When $J_\chi$ is small, we find Bragg peaks at $(\pi,\pi)$ for small $J_2$ and at $(0,\pi)$ for large $J_2$, which are consistent with Neel and stripe magnetic orders, respectively. When $J_\chi$ is large, we find multi-Q feature in spin structure factor, with two peaks at $(0,\pi)$ and $(\pi,0)$, as well as two satellite peaks at $(\pm \pi/2, \pm\pi/2)$, which are consistent with the spin arrangement of non-coplanar chiral spin state~\cite{rabson1995}. As discussed above, these three magnetic orders are also found in the corresponding classical Heisenberg model. In the regime in between these three magnetic ordered phases, no sharp Bragg peak is present, which indicates a nonmagnetic regime where quantum fluctuation is substantially strong to destroy long range orders.
In the lower panel of Fig.~\ref{Fig_structure_factor}, we show the profile of spin structure factors at specific momentum locations $\s(\pi,\pi)$, $\s(\pi,0)$ and $\s(0,\pi)$ in the whole $J_2$-$J_\chi$ phase diagram, where large values are observed in the Neel order phase, stripe order phase, and CSS phase, respectively. They are consistent with the phase boundaries in Fig.~\ref{Fig_phase_diagram}(b), which are also labeled as dashed white lines in Fig.~\ref{Fig_structure_factor} for comparison.

\begin{figure}
    \begin{center}
    \includegraphics[width=0.9\columnwidth]{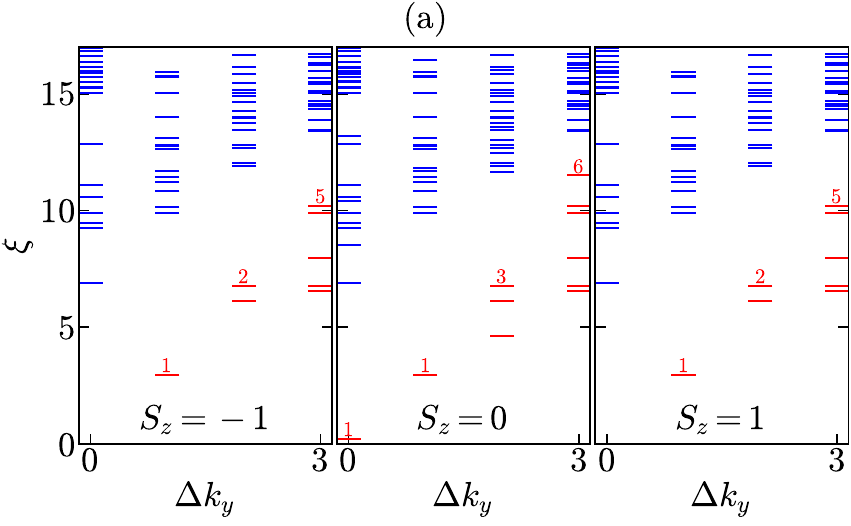}
    \includegraphics[width=0.9\columnwidth]{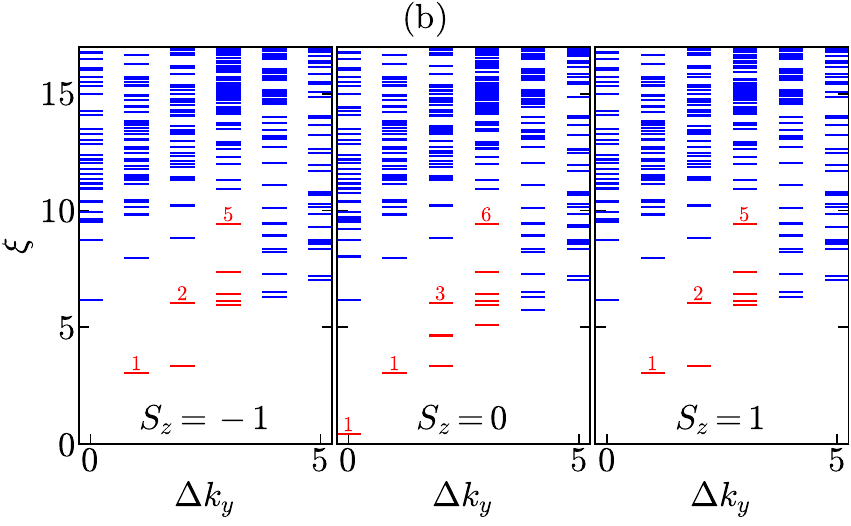}
    \end{center}
\caption{Momentum resolved entanglement spectrum of CSL in (a) $L_y=4$ cylinder and (b) $L_y=6$ cylinder. The low-lying entanglement spectrum clearly shows level counting $\{1,1,3,6,\dots\}$ in $S^z = 0$ sector and $\{1,2,5,\dots\}$ in $S^z = \pm 1$ sectors in both system sizes.}
\label{Fig_spectrum}
\end{figure}

\begin{figure}
    \begin{center}
        \includegraphics[width=0.45\columnwidth]{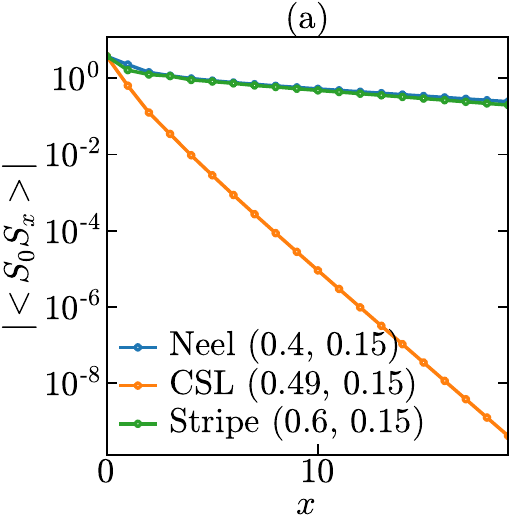}%
        \includegraphics[width=0.45\columnwidth]{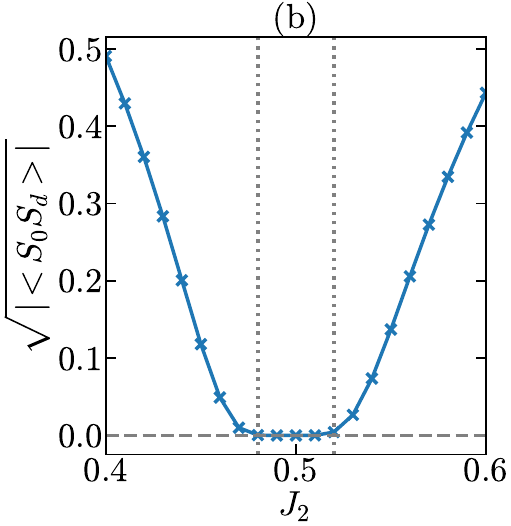}
        \includegraphics[width=0.45\columnwidth]{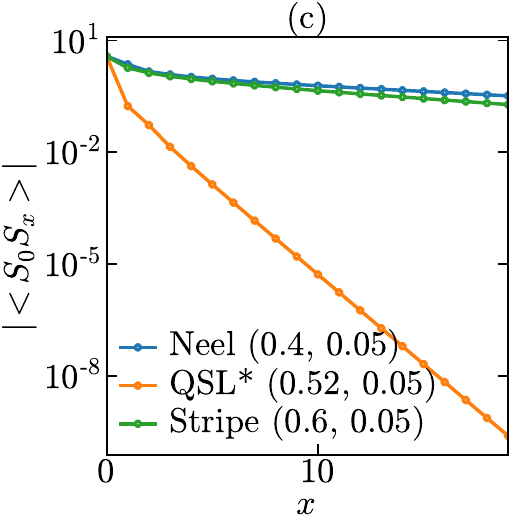}%
        \includegraphics[width=0.45\columnwidth]{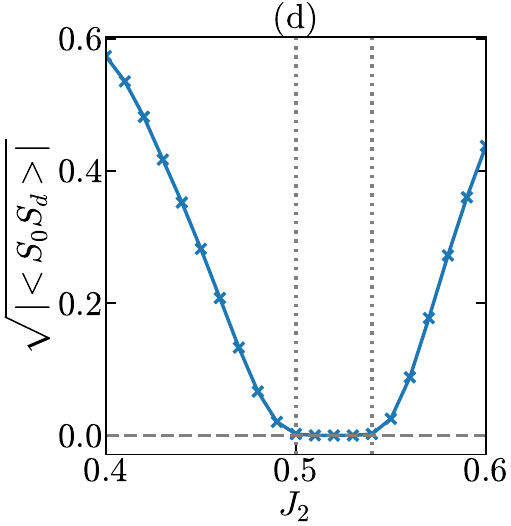}
    \end{center}
    \caption{Spin correlation function $|\langle S_0 S_x\rangle|$ versus distance for different phases at (a) $J_\chi=0.15$ and (c) $J_\chi=0.05$. Long range spin correlation $\sqrt{|\langle S_0 S_d\rangle|}$ versus $J_2$ for (b) $J_\chi=0.15$ and (d) $J_\chi=0.05$. We set $L_y=4$ here. }
    \label{Fig_correlation_Ly4}
\end{figure}

\textit{Read-Rezayi non-Abelian CSL.---}
In the vicinity of the boundaries of magnetic phases, the spin correlation $|\langle \mathbf S_0\cdot  \mathbf S_x\rangle|$ shows an exponential decay along the cylinder, much faster than those in magnetic ordered phases. This is shown in Fig.~\ref{Fig_correlation_Ly4} for $L_y=4$ cylinder (see results for $L_y=6$ cylinder in Supp. Mat. Sec. I ~\cite{SuppMaterial}), where distinct  behaviors of spin correlations are clearly observed in ordered and disordered phases. We have also checked nearest neighbor bond energy in the nonmagnetic regime and  find no lattice translational symmetry  breaking consistent with a uniform state without  valence bond order.

The explicit nature of this underlying spin liquid can be unambiguously revealed by characteristic level countings of edge excitations. Here we extract momentum resolved entanglement spectrum~\cite{li2008,cincio2013} from the matrix product state representation of the ground state, which is found to have one-to-one correspondence with the edge spectrum. In Fig.~\ref{Fig_spectrum}(a), we show the low-lying entanglement spectrum of the ground state at $J_2=0.5$ and $J_\chi=0.15$ in $L_y=4$ cylinder using finite DMRG calculations. We find that the quasi-degenerate level counting exactly matches the tower of states of $SU(2)_3$ Wess-Zumino-Witten theory~\cite{francesco1997} in vacuum sector, which describes edge excitations of the Read-Rezayi fractional quantum Hall state. As anticipated (see Supp. Mat.~\cite{SuppMaterial}), the low-lying entanglement spectrum reveals characteristic level counting of $\{1,1,3,6,\dots\}$ in $S^z = 0$ sector and $\{1,2,5,\dots\}$ in $S^z = \pm 1$ sectors. We also find the same edge countings in $L_y=6$ cylinder, which is shown in Fig.~\ref{Fig_spectrum}(b). Intuitively these characteristic countings can also be understood in terms of generalized Pauli principle~\cite{bernevig2008,zhu2015} in the thin-torus limit, which states no more than $3$ particles in $2$ consecutive orbitals for bosonic Read-Rezayi state indicating the importance of larger local Hilbert space.

\textit{Phase transition.---}
The phase boundaries in the phase diagram can be determined by long-distance spin correlation defined as $\sqrt{|\langle S_0 S_x \rangle|}$, where $x$ denotes the distance between two spins. 
In Fig.~\ref{Fig_correlation_Ly4}(a), for two magnetic phases, the spin correlation decays relatively slow with distances.  They correspond to the Neel order and stripe order, as indicated by the Bragg peaks in spin structure factor in Fig.~\ref{Fig_structure_factor}. 
In Fig.~\ref{Fig_correlation_Ly4}(b), the spin correlation at long distance $d=19$ is vanishing small for $0.48<J_2<0.52, J_\chi=0.15$, leaving a window for the CSL phase. 
A similar behavior is also observed for small values of $J_\chi$. When we vary $J_2$ at fixed $J_\chi=0.05$, we find a similar nonmagnetic phase for $0.5 \le J_2\le 0.54$ as shown in Figs.~\ref{Fig_correlation_Ly4}(c) and (d). In this case, however, the influence of spin chirality term is relatively small and we do not observe characteristic edge countings of the CSL in entanglement spectrum. For this reason, 
we label this quantum disordered regime as QSL*, whose exact nature is beyond current investigation and will be left for future studies. For larger $J_\chi>0.25$ the influence of spin chirality term dominates and stabilizes CSS, which is consistent with classical analysis. The quantum phase diagram shown in Fig.~\ref{Fig_phase_diagram} is obtained in $L_y=4$ system. We have also checked the phase diagram for $L_y=6$ and observe a similar transition behavior (see Supp. Mat.~\cite{SuppMaterial}) We find all the quantum phases found in $L_y=4$ cylinder persists for the $L_y=6$ system.

\textit{Summary and discussion.---}
We have numerically studied the spin-$3/2$ $J_1$-$J_2$-$J_\chi$ Heisenberg model on the square lattice using unbiased density-matrix renormalization group calculation. We map out the global quantum phase diagram that contains three conventional magnetic orders, including a Neel state, a stripe state, and a chiral spin state. Crucially, among these magnetic phases an interesting chiral spin liquid (CSL) is uncovered, where the quantum fluctuations are strongly enhanced near the phase boundaries of different magnetic ordered phases. The quantum fluctuations destroy conventional long-ranged magnetic orders and induce a CSL. The nature of this CSL is identified as a non-Abelian Read-Rezayi state via characteristic  momentum resolved entanglement spectrum. Additionally, we also find another possible quantum spin liquid near $J_\chi\sim 0$, whose exact nature deserves future study.
Our findings demonstrate the existence of non-Abelian CSL in higher spin quantum antiferromagnets which supports the non-Abelian Fibonacci anyonic statistics and also paves the way to searching other intriguing QSLs in these higher spin systems 
via the mechanism of enhancing quantum fluctuations through tuning competing interactions. 
Our model Hamiltonian may be realized  in  the 3d transition metal compound $\text{Ba}_{2}\text{Co}\text{Ge}_{2}\text{O}_{7}$~\cite{miyahara2011,romhanyi2012} with effective spin-3/2 antiferromagnetic Heisenberg exchange, where the effective chiral spin interactions can be induced by an applied out-of-plane magnetic field.

At last, several remarks are given in order.
First,  in comparison with the non-Abelian spin liquid with Ising-type anyons in $S=1$ model~\cite{huang2022}, the spin liquid in spin-$3/2$ model is not only a simple extension,  but a new class of exotic topological orders which will inspire and call for new theoretical proposals.
Another difference is that Fibonacci anyon does not have a free-Fermion description, in sharp contrast to the Ising anyon~\cite{kitaev2001}.
Second,  how to understand the current findings in spin-$3/2$ model and connect the current model with previous studies in soft-boson model~\cite{cooper2001} are highly nontrival. One possible way to think about it is from  coupled-wire construction~\cite{kane2002,teo2014} or the projective construction~\cite{cappelli2001,regnault2008,barkeshli2010b,vaezi2014}, i.e. 
if treating a spin-$3/2$ system as three coupled spin-$1/2$ layers, one could image the non-Abelian $Z_3$ order is manifested by three Abelian spin liquid states.  However, the effective interactions which allow such a state to emerge from three coupled spin-1/2 systems can become another challenge issue demanding future studies.

%\begin{acknowledgments}
W.Z. thanks Z. X. Liu for simulating discussions. This work was supported by National Science Foundation of China under project number 92165102, 11974288 (W.W.L., W.Z.).
Y.H. was supported by the U.S. DOE NNSA under Contract No. 89233218CNA000001 and by the Center for Integrated Nanotechnologies, a DOE BES user facility, in partnership with the LANL Institutional Computing Program for computational resources.
D.N.S. was mainly supported by  National Science Foundation (NSF) Partnership in Research and Education in Materials  DMR-1828019 and  partially supported by  NSF   Princeton Center for Complex Materials, a Materials Research Science and Engineering Center DMR-2011750. 
% \end{acknowledgments}

\bibliography{J1J2JxSQ}

\clearpage

\newcommand{\beginsupplement}{%
        \setcounter{table}{0}
        \renewcommand{\thetable}{S\arabic{table}}%
        \setcounter{figure}{0}
        \renewcommand{\thefigure}{S\arabic{figure}}%
        \setcounter{section}{0}
        \renewcommand{\thesection}{\Roman{section}}%
        \setcounter{equation}{0}
        \renewcommand{\theequation}{S\arabic{equation}}%
        }
\clearpage
\onecolumngrid
\beginsupplement

\begin{center}
	\textbf{\large Supplementary Materials for `` Global quantum phase diagram and non-Abelian chiral spin liquid in a spin-3/2 square lattice antiferromagnet ''}
\end{center}
	\vspace{4mm}

In this supplementary materials, we present more details to support the conclusion in the main text. Sec.~\ref{Spin correlations} shows the spin correlations as a function of bond dimensions, as a consistent check of the robustness of numerical results. 
We also show the spin correlations on $L_y=6$
cylinder. 
Sec.~\ref{Dimer correlations} presents the dimer correlations and discusses the possibility of valence bond solids. 
In Sec.~\ref{Phase transition}, we show the evidence of phase transitions on a wider cylinder $L_y=6$. 
In Sec.~\ref{Edge counting}, we list the countings of the effect edge theory of the bosonic Read-Rezayi state.
In Sec.~\ref{nematicity}, we examine the possible nematicity in the QSL*.

\section{Spin correlations }
\label{Spin correlations}

\begin{figure}[b]
    \begin{center}
        \includegraphics[width=0.8\columnwidth]{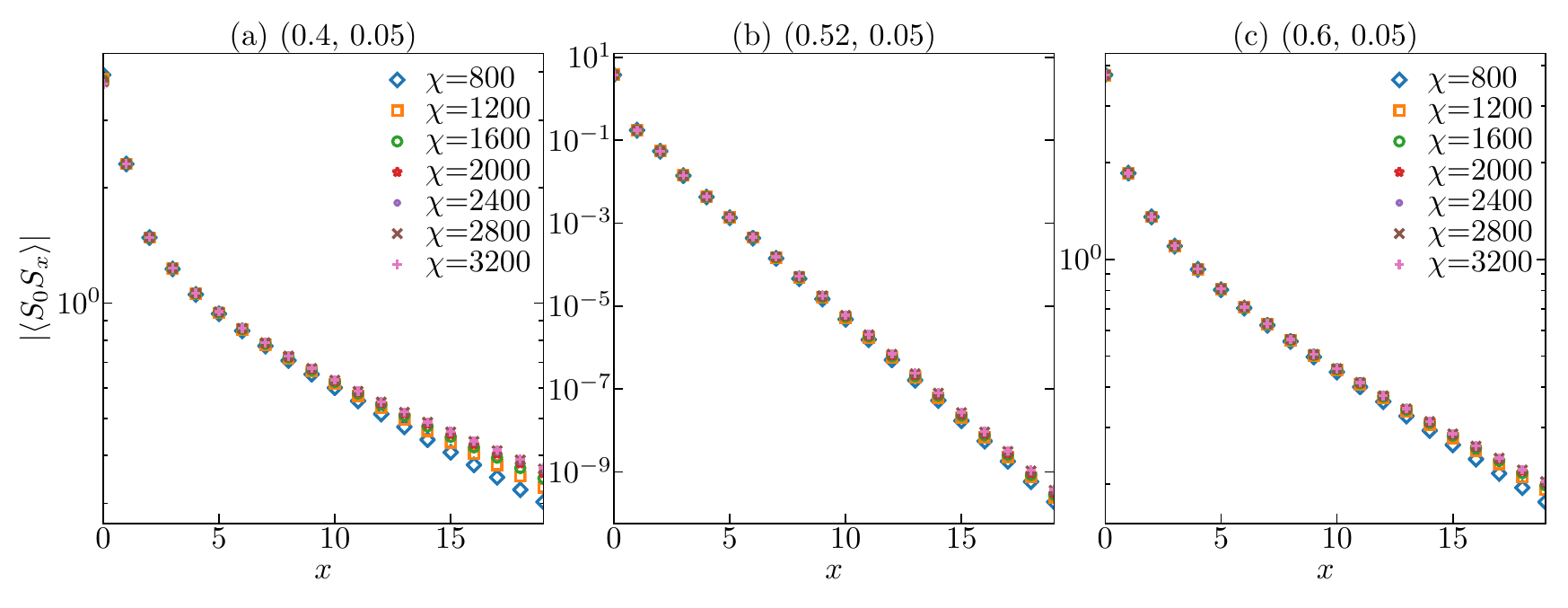}\\
        \includegraphics[width=0.8\columnwidth]{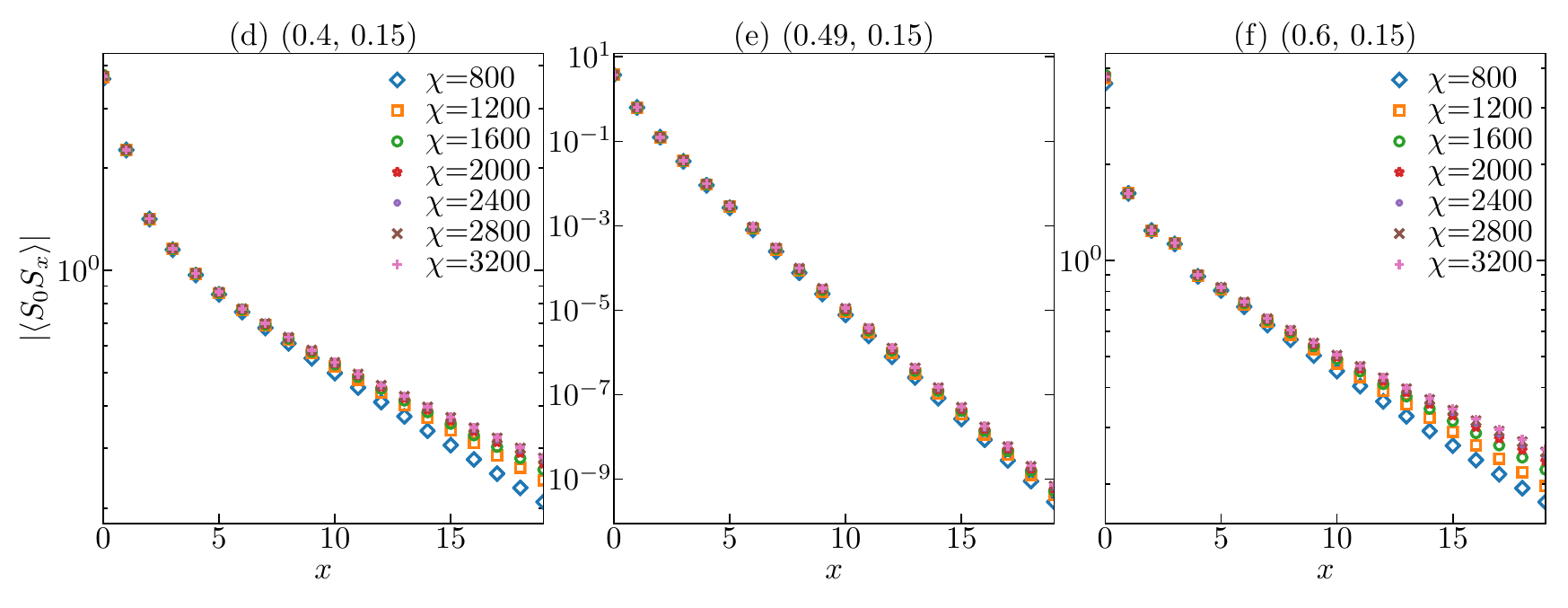}
    \end{center}
    \caption{Spin correlation function $|\langle \mathbf S_0 \cdot \mathbf S_x\rangle|$ versus distance for various bond dimensions. Here we set $L=4$, $J_\chi=0.05$ (upper panel) and $J_\chi=0.15$ (lowere panel). Please note that the scale in (b) and (e) are different from the others.  }
    \label{Fig_corr_005}
\end{figure}

In Fig.~\ref{Fig_corr_005} we show the behavior of spin correlation in different phases on $L_y=4$ cylinder. In this case spin correlations has ignorable dependence on the maximum bond dimension used in DMRG calculations, indicating well converged behavior. For the QSL* state at $(0.52,0.05)$ and CSL state at $(0.49,0.15)$, we use semi-log plot and clearly find exponentially decaying behavior in all bond dimensions. On the other hand,  the spin correlations in magnetic phases decay much slower than those in spin liquid phases, and tends to a power-law behavior in large bond dimension limit.\\

\begin{figure}
    \begin{center}
        \includegraphics[width=0.8\columnwidth]{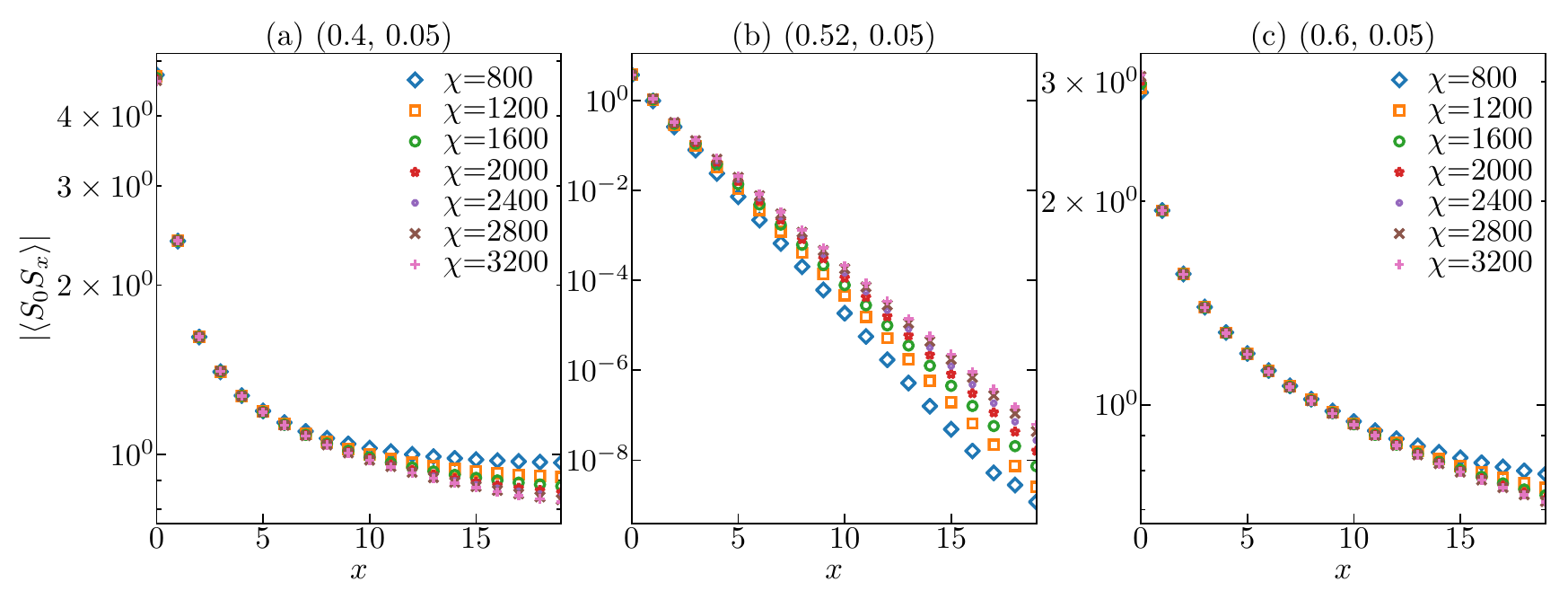}\\
        \includegraphics[width=0.8\columnwidth]{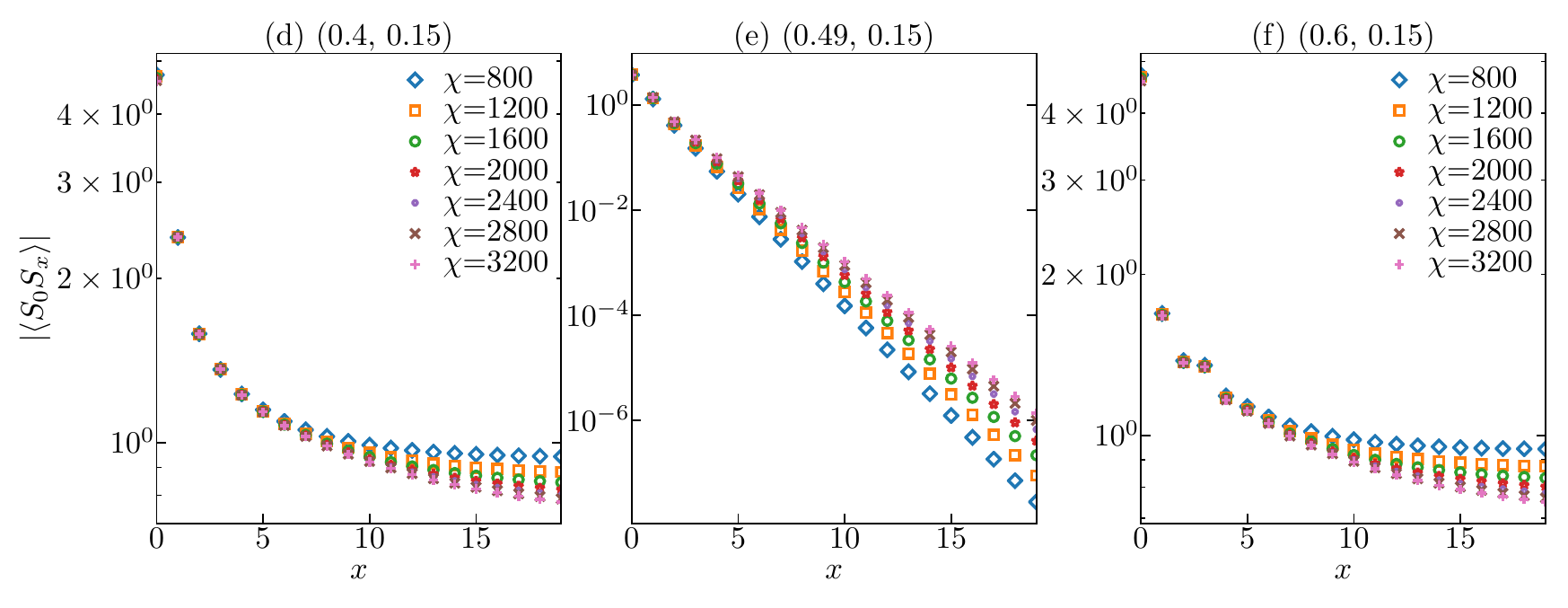}
    \end{center}
    \caption{Spin correlation function $|\langle \mathbf S_0 \cdot \mathbf S_x\rangle|$ versus distance for various bond dimensions. Here we set $L=6$, $J_\chi=0.05$ (upper panel) and $J_\chi=0.15$ (lowere panel).  Please note that the scale in (b) and (e) are different from the others. }
    \label{Fig_corr_015}
\end{figure}

In Fig.~\ref{Fig_corr_015} we show the behavior of spin correlation in different phases on $L_y=6$ cylinder. As in the case of $L_y=4$ cylinder, we also find exponential spin decay in QSL* and CSL phases for all studied bond dimensions. On the other magnetic phases, the spin correlations decays much slower as expected. On increasing the bond dimension, we find the tendency of power-law decay in these magnetic phases.\\

\section{Dimer correlations}
\label{Dimer correlations}

We examine the dimer-dimer correlation function $\langle D^x_0 D^x_x \rangle$ in QSL* phase to see whether valence bond solid may exists. The dimer operator is defined by 
\begin{equation}
D^\alpha_x =   \mathbf S_x \cdot \mathbf S_{x+\alpha}
\end{equation}
where $\alpha = \hat x (\hat y)$ labels the nearest neighbor site along $x (y)$ direction.

\begin{figure}
    \begin{center}
        \includegraphics[width=0.5\columnwidth]{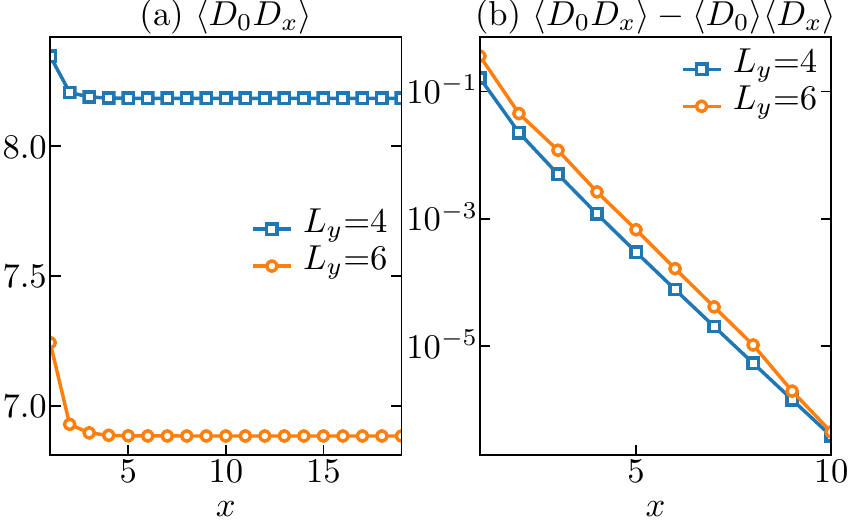}    \end{center}
    \caption{Dimer correlation function $\langle D_0 D_x \rangle$ versus distance for different phases at (0.52,0.05)}
    \label{Fig_dimer_correlation}
\end{figure}

As shown in Fig.~\ref{Fig_dimer_correlation},
we find a uniform dimer-dimer correlation in both $L_y=4$ cylinder and $L_y=6$ cylinder. No signal of ``strong-weak" pattern is found. Moreover, we also show the fluctuation term $\langle D^x_0 D^x_x \rangle- \langle D_0 \rangle \langle D_x \rangle$, we observe an exponentially decaying behavior. And the correlation length does not increase if going from $L_y=4$ to $L_y=6$. These observations rule out the possibility of valence bond order. \\

\section{Phase transition on $L_y=6$ cylinder}
\label{Phase transition}

\begin{figure}
    \begin{center}
        \includegraphics[width=0.25\columnwidth]{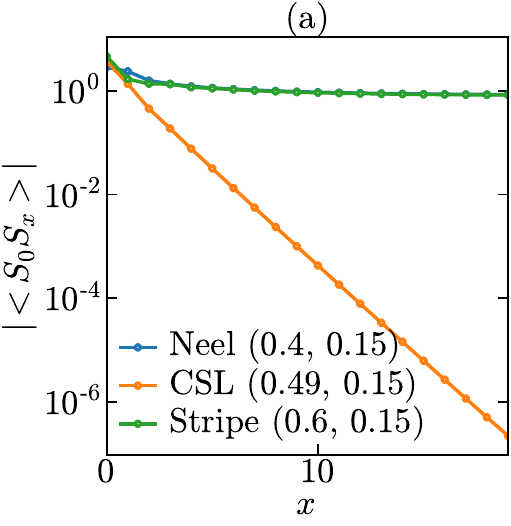}%
        \includegraphics[width=0.25\columnwidth]{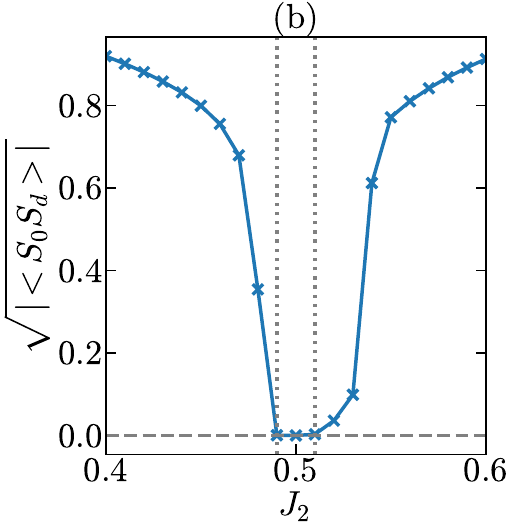}
        \includegraphics[width=0.25\columnwidth]{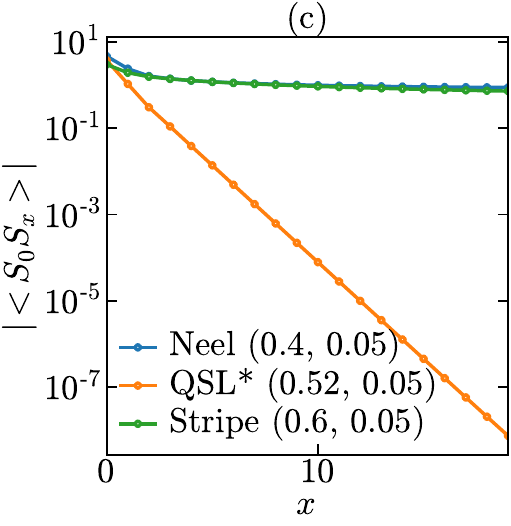}%
        \includegraphics[width=0.25\columnwidth]{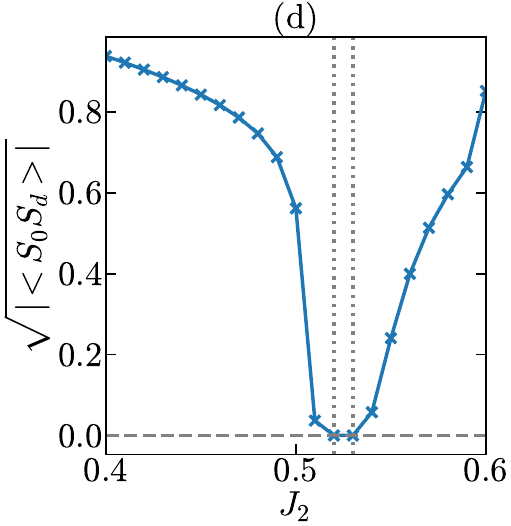}
    \end{center}
    \caption{Spin correlation function $|\langle \mathbf S_0 \cdot \mathbf S_x\rangle|$ versus distance for different phases at (a) $J_\chi=0.15$ and (c) $J_\chi=0.05$. Long range spin correlation $\sqrt{|\langle S_0 S_d\rangle|}$ versus $J_2$ for (b) $J_\chi=0.15$ and (d) $J_\chi=0.05$. We set $L_y=6$ here.}
    \label{Fig_correlation_Ly6}
\end{figure}

We show phase transitions in $L_y=6$ cylinder at fixed $J_\chi=0.15$ and $J_\chi=0.05$ in Fig.~\ref{Fig_correlation_Ly6}. (In Fig. 4 in main text, we show the results on $L_y=4$.) In both cases, we find Neel order at small $J_2$ and stripe order at large $J_2$, both of which possess long range magnetic order. At intermediate $J_2$ a disordered regime exists, which correspond to CSL at $J_\chi=0.15$ and QSL* phase at $J_\chi=0.05$. 
Based on the calculations on $L_y=4,6$, we believe the finding of CSL and QSL* in this $J_1$-$J_2$-$J_{\chi}$ model is robust. Nevertheless, we cannot totally exclude the finite-size effect beyond $L_y>6$, which is out of reach of our computational capbility. 

\section{Edge counting of Read-Rezayi state}
\label{Edge counting}
Here we list degeneracy sequences of edge modes in all topological sectors of $\nu=3/2$ bosonic Read-Rezayi (RR) state. For the details one may consult the references like \cite{ardonne2001,Ardonne2004,haldane2007,bernevig2008}. We didnot find these information and results in the literature, so we list them here.  

Starting from a highest density root configuration with momentum $\Delta L=0$, we can enumerate all admissible configurations constrained by generalized Pauli principle, which states a fractional exclusion statistics that no more than 3 bosons are allowed in 2 consecutive orbitals in this RR state. These admissible configurations are in one to one correspondence with RR edge modes and thus yields the degeneracy in each momentum sector. The following tables list all possible countings for $\Delta L\le 4$. They correspond to two root configurations ...303030303... and ...2121212121... with different number of bosons.

\begin{table}[h]
    \caption{bosonic RR state root configuration ...303030303 (3n) with edge counting 1,1,3,6,12}
\begin{tabular}{ccccc}
    \hline\hline
    $\Delta L=0$ & $\Delta L=1$ & $\Delta L=2$ & $\Delta L=3$ & $\Delta L=4$\\
303030303 & 3030303021 & 30303030201 & 303030302001 & 3030303020001 \\
  &   & 30303021210 & 303030212010 & 3030302120010 \\
  &   & 30303030120 & 303021212100 & 3030212120100 \\
  &   &   & 303030211200 & 3021212121000 \\
  &   &   & 303030301110 & 3030212112000 \\
  &   &   & 303030300300 & 3030302030100 \\
  &   &   &   & 3030302111100 \\
  &   &   &   & 3030301212000 \\
  &   &   &   & 3030302103000 \\
  &   &   &   & 3030303011010 \\
  &   &   &   & 3030303002100 \\
  &   &   &   & 3030303010200 \\
    \hline
    \hline\hline
\end{tabular}
\end{table}

\begin{table}[h]
    \caption{bosonic RR state root configuration ...303030302 (3n+2) or ...303030301 (3n+1) with edge counting 1,2,5,9,18}
\begin{tabular}{ccccc}
    \hline\hline
    $\Delta L=0$ & $\Delta L=1$ & $\Delta L=2$ & $\Delta L=3$ & $\Delta L=4$\\
303030302 & 3030303011 & 30303030101 & 303030301001 & 3030303010001 \\
  & 3030302120 & 30303021110 & 303030211010 & 3030302110010 \\
  &   & 30302121200 & 303021211100 & 3030212110100 \\
  &   & 30303020300 & 302121212000 & 3021212111000 \\
  &   & 30303030020 & 303021203000 & 2121212120000 \\
  &   &   & 303030121100 & 3021212030000 \\
  &   &   & 303030202100 & 3030211211000 \\
  &   &   & 303030210200 & 3030212021000 \\
  &   &   & 303030300110 & 3030203030000 \\
  &   &   &   & 3030212102000 \\
  &   &   &   & 3030301210100 \\
  &   &   &   & 3030301121000 \\
  &   &   &   & 3030301202000 \\
  &   &   &   & 3030302020100 \\
  &   &   &   & 3030302012000 \\
  &   &   &   & 3030302101100 \\
  &   &   &   & 3030303001010 \\
  &   &   &   & 3030303000200 \\
    \hline
    \hline\hline
\end{tabular}
\end{table}

\begin{table}[h]
    \caption{bosonic RR state root configuration ...2121212121 (3n) or ...212121212 (3n+2) with edge counting 1,2,5,10,20}
\begin{tabular}{ccccc}
    \hline\hline
    $\Delta L=0$ & $\Delta L=1$ & $\Delta L=2$ & $\Delta L=3$ & $\Delta L=4$\\
2121212121 & 21212121201 & 212121212001 & 2121212120001 & 21212121200001 \\
  & 21212121120 & 212121203010 & 2121212030010 & 21212120300010 \\
  &   & 212121211110 & 2121203030100 & 21212030300100 \\
  &   & 212121121200 & 2121212021100 & 21203030301000 \\
  &   & 212121210300 & 2121211211100 & 21212030211000 \\
  &   &   & 2121121212000 & 21212120210100 \\
  &   &   & 2121211203000 & 21212112110100 \\
  &   &   & 2121212111010 & 21211212111000 \\
  &   &   & 2121212102100 & 21121212120000 \\
  &   &   & 2121212110200 & 21211212030000 \\
  &   &   &   & 21212111211000 \\
  &   &   &   & 21212112021000 \\
  &   &   &   & 21212103030000 \\
  &   &   &   & 21212112102000 \\
  &   &   &   & 21212120121000 \\
  &   &   &   & 21212120202000 \\
  &   &   &   & 21212121110010 \\
  &   &   &   & 21212121020100 \\
  &   &   &   & 21212121012000 \\
  &   &   &   & 21212121101100 \\
    \hline
    \hline\hline
\end{tabular}
\end{table}

\begin{table}[h]
    \caption{bosonic RR state root configuration ...212121211 (3n+1) with edge counting 1,3,6,13,24}
\begin{tabular}{ccccc}
    \hline\hline
    $\Delta L=0$ & $\Delta L=1$ & $\Delta L=2$ & $\Delta L=3$ & $\Delta L=4$\\
    \hline
212121211 & 2121212101 & 21212121001 & 212121210001 & 2121212100001 \\
  & 2121211210 & 21212112010 & 212121120010 & 2121211200010 \\
  & 2121212020 & 21211212100 & 212112120100 & 2121121200100 \\
  &   & 21212111200 & 211212121000 & 2112121201000 \\
  &   & 21212030200 & 212112112000 & 1212121210000 \\
  &   & 21212120110 & 212121030100 & 2112121120000 \\
  &   &   & 212121111100 & 2121120301000 \\
  &   &   & 212120301100 & 2121121111000 \\
  &   &   & 212030302000 & 2121112120000 \\
  &   &   & 212120212000 & 2120302120000 \\
  &   &   & 212121103000 & 2030303020000 \\
  &   &   & 212121201010 & 2121121030000 \\
  &   &   & 212121200200 & 2121210300100 \\
  &   &   &   & 2121210211000 \\
  &   &   &   & 2121202111000 \\
  &   &   &   & 2120303011000 \\
  &   &   &   & 2121202030000 \\
  &   &   &   & 2121211110100 \\
  &   &   &   & 2121203010100 \\
  &   &   &   & 2121203002000 \\
  &   &   &   & 2121211021000 \\
  &   &   &   & 2121211102000 \\
  &   &   &   & 2121212010010 \\
  &   &   &   & 2121212001100 \\
    \hline\hline
\end{tabular}
\end{table}

\section{Possible nematicity in spin liquid phases }
\label{nematicity}

\begin{figure}
    \begin{center}
        \includegraphics[width=0.33\columnwidth]{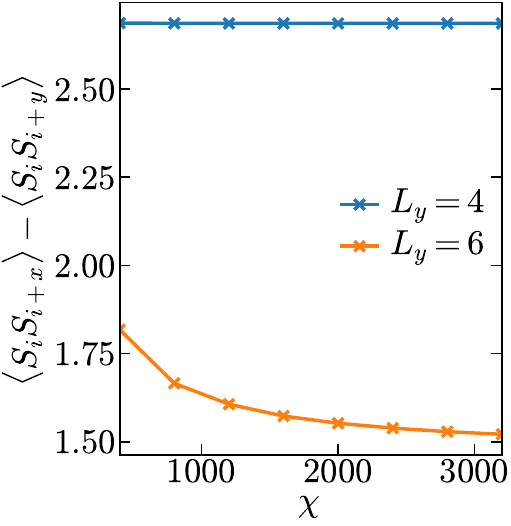}%
        \includegraphics[width=0.66\columnwidth]{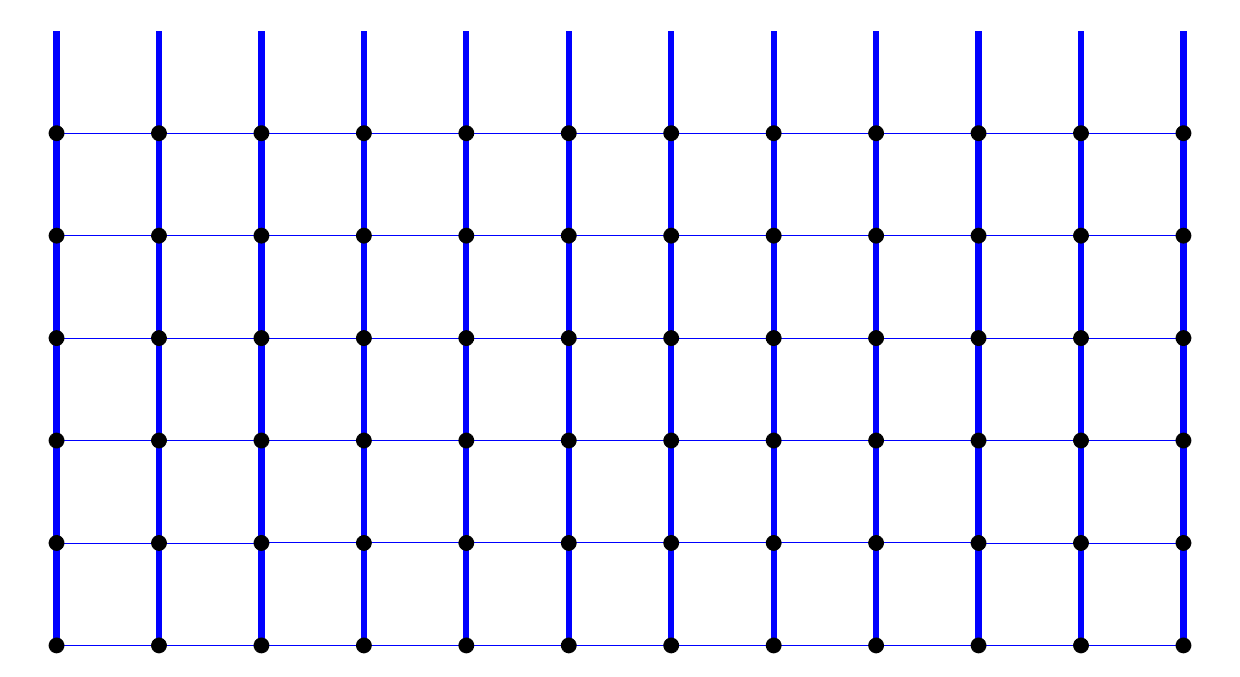}
    \end{center}
    \caption{Nematicity in QSL* on increasing bond dimension $\chi$ for different cylinder circumferences $L_y$. We also show the real-space distribution of NN bond energy on the right side, which corresponds to QSL* phase at (0.52,0.05) on a $12\times 6$ finite cylinder. The line width and color (red/blue for positive/negative) denote the amplitude and sign of corresponding bond energy.  }
    \label{Fig_nematicity}
\end{figure}

In our performed DMRG calculations, we notice possisble nematicity in QSL* phases. The nematic order can be measured by the difference of bond energies along $x$ and $y$ directions as $\langle S_i S_{i+x}\rangle-\langle S_i S_{i+y}\rangle$, where $i$ labels lattice sites in the bulk. Fig.~\ref{Fig_nematicity} shows the tendency of nematic order on increasing bond dimension $\chi$ or cylinder circumference $L_y$. Although the nematicity is quite strong in each case, it quickly reduces by increasing cylinder width from $L_y=4$ to $L_y=6$. Due to the limited system sizes accessible, we cannot determine whether such nematic order persists in the thermodynamic limit.
If this nematic order persists in larger system size, this indicates the QSL* phase has a rotational symmetry breaking nature. Since available system sizes is limited in DMRG calculation, we propose to study this possible QSL* phase using other complementary methods such as variational Monte Carlo.

\section{Entanglement spectrum in QSL*}

\begin{figure}
    \begin{center}
        \includegraphics[width=0.6\columnwidth]{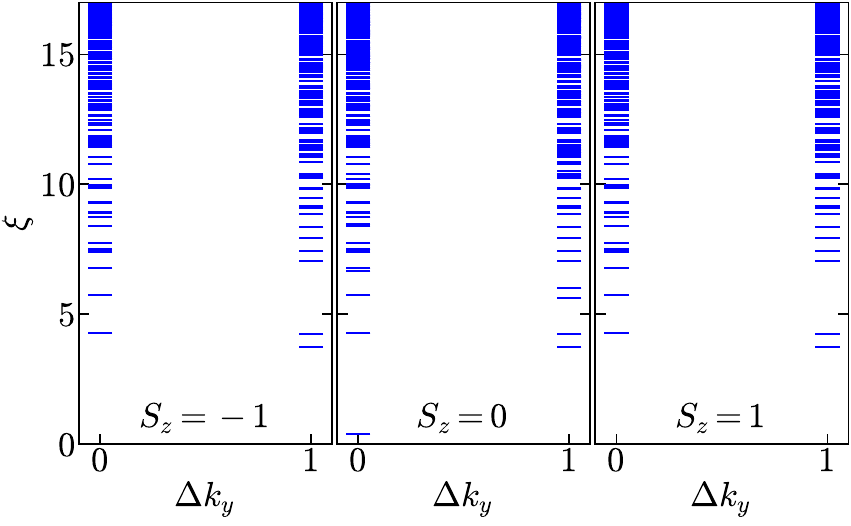}
    \end{center}
    \caption{Entanglement spectrum in the QSL* phase at $(0.52, 0)$ on a $L_y=6$ cylinder.}
    \label{Fig_QSL_ES}
\end{figure}

As is shown in the main text, the entanglement spectrum of CSL state exhibits characteristic quasi-degenerate pattern that manifests its underlying topological order. On the contrary, the QSL* found at small $J_\chi$ does not possess such clear signature. We show in Fig.~\ref{Fig_QSL_ES} the entanglement spectrum of disordered QSL* at $(0.52, 0)$. The entanglement spectrum in this extreme case has only two possible momentum quantum numbers due to the real-valued wave function. We can see clear distinctions between the disordered CSL and QSL* phases.

\end{document}